\documentclass[12pt]{article}
\usepackage{amssymb}
\usepackage{amsmath}
\usepackage{IEEEtrantools}
\usepackage{pdflscape}
\usepackage{afterpage}
\usepackage{capt-of}

\usepackage{graphicx}
\usepackage{epstopdf}
\usepackage{subfig}

\usepackage{amsfonts,amsthm}
\usepackage{amsmath}  
\usepackage{latexsym}
\usepackage{amssymb}
\usepackage{mathrsfs}
\usepackage{tikz}
\usepackage[all]{xy}
\usepackage[normalem]{ulem}
\usepackage{hyperref}

\usepackage{changepage}

\numberwithin{equation}{section}

\newcommand{\be}{\begin{equation}}
\newcommand{\ee}{\end{equation}}
\newcommand{\bea}{\begin{eqnarray}}
\newcommand{\eea}{\end{eqnarray}}
\newcommand{\non}{\nonumber}

\newcommand{\C}{\mathbb{C}}
\newcommand{\tr}{\mathop{\rm tr}\nolimits}
\newcommand{\diag}{\mathop{\rm diag}\nolimits}

\newcommand{\nicolas}[1]{$\framebox{\tiny n}$\ \textbf{\texttt{{\color{blue}\footnotesize#1}}}}

\begin{document}

\begin{titlepage}
\strut\hfill UMTG--296
\vspace{.5in}

\begin{center}

\LARGE Equivalent T-Q relations and exact results\\[0.2in] 
for the open TASEP\\
\vspace{1in}
\large 
Nicolas Crampe \footnote{Laboratoire Charles Coulomb (L2C), Univ Montpellier, CNRS, Montpellier, France\\nicolas.crampe@umontpellier.fr} and
Rafael I. Nepomechie \footnote{Physics Department,
P.O. Box 248046, University of Miami, Coral Gables, FL 33124 USA\\ nepomechie@miami.edu}
\\[0.8in]
\end{center}

\vspace{.5in}

\begin{abstract}
 Starting from the Bethe ansatz solution for the open Totally 
 Asymmetric Simple Exclusion Process (TASEP), we compute the largest eigenvalue of the deformed Markovian 
 matrix, in exact agreement with results obtained by the matrix ansatz.
 We also compute the eigenvalues of the higher conserved 
 charges.
 The key step is to find a simpler equivalent T-Q relation,
 which is similar to the one for the TASEP with periodic boundary 
 conditions.
\end{abstract}

\end{titlepage}

\setcounter{footnote}{0}

\section{Introduction}\label{sec:intro}

Numerous analytical results have been obtained in
non-equilibrium statistical mechanics by using methods
that were originally 
developed for quantum
integrable systems.  Indeed, a connection between the Markovian matrix
of the Asymmetric Simple Exclusion Process (ASEP) and
a Hamiltonian describing an integrable quantum spin
chain has been pointed out in \cite{San}.  The Bethe ansatz, which was
invented to compute the spectrum of quantum models, can therefore also be used in
the context of exclusion processes.  For example,
the Bethe ansatz was used to compute the
current fluctuations \cite{DL, PM, Pro, RS} and their
spectral gaps \cite{dE,Pro17} for various exclusion
processes. 

Another method, called the matrix ansatz or DEHP method \cite{DEHP}, 
was specially developed to compute the stationary states
as well as various correlation functions.
This approach was
subsequently used to compute the fluctuations of the current
for the Totally ASEP (TASEP) with open boundaries \cite{LM} and for
the ASEP \cite{GLMV}.  The Bethe ansatz was not used for such
computations since the necessary generalization of the Bethe ansatz
for generic boundary conditions
had not been developed until recently
\cite{Cao:2013qxa, Nepomechie:2013ila, BC, KMN, Wang2015, BP, C17}.
Based on these generalizations, the Bethe ansatz for the
TASEP with open boundary conditions was formulated in \cite{C14}, and for the
ASEP in \cite{WYCCY}. The main purpose of this letter
is to use the recent Bethe ansatz results to recover the analytical results obtained by matrix
ansatz. To this end, we map the so-called T-Q relation obtained in \cite{C14} for the
open TASEP to a simpler equivalent T-Q relation, which is similar to the one for the TASEP
with {\em periodic} boundary conditions. Then, using the same approach
developed in \cite{DL}, we recover the matrix ansatz results of \cite{LM}. 
Remarkably, the two distinct T-Q relations describe exactly the 
same transfer-matrix  eigenvalue (``T''), for any value of length $L$. It would be very
interesting to find more such examples.

The plan of the paper is as follows.  In Section \ref{sec:model}
we recall the definition of the open TASEP model, and its
one-parameter ($\mu$) deformation (deformed Markovian
matrix), which allows one to compute the fluctuations of the current.  The
transfer matrix associated to this model as well as its eigenvalues
expressed in terms of Bethe roots (i.e., the T-Q relation)
are briefly reviewed in Section
\ref{sec:transfer}.  A perturbative approach for computing 
the eigenvalue for 
small values of $\mu$ is presented in Section \ref{sec:smmu}.  The
main result of this paper is presented in Section \ref{sec:eqTQ}, where 
a second T-Q relation is introduced, which reproduces the largest eigenvalue of the
deformed Markovian matrix.  In Section \ref{sec:ex}, we show how 
the matrix ansatz results of \cite{LM} can be recovered and 
generalized from these new Bethe equations.

\section{Bethe ansatz for the open TASEP}

\subsection{The open TASEP model\label{sec:model}}

\begin{figure}[htb]
\begin{center}
\begin{tikzpicture}[scale=0.5]
\fill (5,1) circle (0.5cm);\fill (7,1) circle (0.5cm);\fill (25,1) circle (0.5cm);\fill (13,1) circle (0.5cm);\fill (-2,2.5) circle (0.5cm);
\draw[thick] (0,0) -- (26,0);
\foreach \x in {0,2,4,...,26} \draw[thick] (\x,0) -- (\x,1);
%\draw[thick,->] (5.4,2)..controls +(0.5,1) and +(-0.5,1) .. (6.6,2);
\draw[thick,->] (7.4,2)..controls +(0.5,1) and +(-0.5,1) .. (8.6,2);\node at (8,3.5) {$1$};
\draw[thick,->] (13.4,2)..controls +(0.5,1) and +(-0.5,1) .. (14.6,2);\node at (14,3.5) {$1$};
\draw[thick,->] (-1,2.5)..controls +(0.5,0) and +(-0.5,1) .. (0.6,1.5);\node at (0,3) {$\alpha$};
\draw[thick,->] (25.4,2)..controls +(0.5,1) and +(-0.5,0) .. (27.5,3);\node at (26,3.5) {$\beta$};
\end{tikzpicture}
\caption{Dynamical rules for the TASEP\label{fig:TASEP}}
\end{center}
\end{figure}
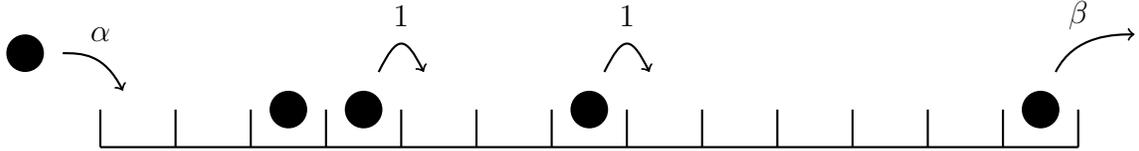

The Totally Asymmetric Simple Exclusion Process (TASEP) on a segment
of $L$ sites in contact with particle reservoirs at the two 
ends
is a stochastic model, which 
evolves in time according to the following rules: during any 
time interval $dt$, each particle jumps
with probability $dt$ to 
the neighboring site on its right if that site is empty,
enters at site $1$ (if no particle is present there) with probability
$\alpha\, dt$, or exits from site $L$ with probability 
$\beta\, dt$. This process is ``totally asymmetric,'' since 
particles cannot jump to the left; ``simple'', since particles can 
jump no more than 1 site in the time $dt$; and ``exclusive'', since a 
site cannot be 
occupied by more than one particle. A schematic representation of the dynamical rules is displayed in
Figure \ref{fig:TASEP}.

The probabilities $P_t( C )$ of finding the system in the configuration $C$ at time $t$ satisfy
\begin{equation}
 \frac{d}{dt} P_t(C) = \sum_{C'} M(C,C') P_t(C')\;,
\end{equation}
where, for $C'\neq C$, $M(C,C')dt$ is the probability to go from $C'$ to $C$ during the interval $dt$ and $M(C,C)=-\sum_{C'\neq C} M(C',C)$.
In a suitable basis, the Markovian matrix for the TASEP is a matrix acting on $(\C^2)^{\otimes L}$
\begin{equation}
 M=B_1 +\sum_{k=1}^{L-1}  w_{k,k+1} +\overline B_L\;,
 \label{Markovian}
\end{equation}
where the subscripts indicate on which space the following matrices 
$w$, $B$ and $\overline B$ act non trivially
\begin{equation}
 B=\begin{pmatrix}
    -\alpha& 0\\
    \alpha & 0
   \end{pmatrix}
\quad , \qquad
w=\begin{pmatrix}
   0 & 0 & 0 & 0\\
   0 & 0 & 1 & 0\\
   0 & 0 & -1& 0\\
   0 & 0 & 0 & 0
  \end{pmatrix}
  \quad \text{and} \qquad
 \overline B=\begin{pmatrix}
   0 & \beta\\
   0 & -\beta
   \end{pmatrix} \;.
\end{equation}
We are in fact more interested in a deformation of the Markovian matrix (\ref{Markovian})
\begin{equation}
M(\mu)=B_1(\mu) +\sum_{k=1}^{L-1}  w_{k,k+1} +\overline B_L\quad\text{where}\qquad B(\mu)=\begin{pmatrix}
    -\alpha& 0\\
    \alpha e^\mu & 0
   \end{pmatrix}\;,\label{eq:Mmu}
\end{equation}
and $\mu$ is a real parameter. Evidently, $M(\mu) \rightarrow 
M$ as $\mu \rightarrow 0$.
This deformed Markovian matrix is important, since its largest
eigenvalue $\lambda(\mu)$ can be used to determine
the fluctuations of the
current entering in the system \cite{LS}.  The latter has been computed
previously in \cite{LM} by using a generalization of the matrix
ansatz \cite{DEHP}. We shall
recover this result (and generalize it) by using the Bethe 
ansatz solution
discovered recently in \cite{C14} following the lines of \cite{DL},
where similar computations have been done for a periodic TASEP.

\subsection{The T-Q relation for the open TASEP 
\label{sec:transfer}}

As usual in the context of the algebraic Bethe ansatz, instead of
diagonalizing only the Markovian matrix, we diagonalize a transfer
matrix depending on a parameter $x$ (called the spectral parameter) from
which we can recover the (deformed) Markovian matrix.  The transfer matrix associated
to the deformed Markovian matrix \eqref{eq:Mmu} is given by \cite{C14}
\begin{equation}
 t(x)=tr_0\big(\ \widetilde{K}_0(x)\ R_{0L}(x)\dots R_{01}(x) \ K_0(x)\ R_{10}(x) \dots R_{L0}(x)\ \big)\;,
 \label{transfer}
\end{equation}
where
\begin{equation}
\widetilde K(x)=\frac{1}{1+xb}\begin{pmatrix}
                 1 & 1\\
                 0 &xb
                \end{pmatrix}
\ , \quad
 R(x)=\begin{pmatrix}
       1 & 0 & 0 & 0\\
       0 & 0 & x & 0\\
       0 & 1 & 1-x & 0\\
       0 & 0 & 0 & 1
      \end{pmatrix}
      \ \text{and}\quad
K(x)=\begin{pmatrix}
      \frac{(a+x)x}{xa+1}& 0\\
      e^\mu \frac{1-x^2}{xa+1} & 1
     \end{pmatrix}\,,
     \label{RKmats}
\end{equation}
and the parameters $a$ and $b$ above are related to 
the boundary rates $\alpha$ and $\beta$ as follows
\begin{equation}
    a=\frac{1}{\alpha}-1\quad\text{and} \qquad b=\frac{1}{\beta}-1\;.
\end{equation}
A key feature of the transfer 
matrix is its commutativity property
\begin{equation}
[t(x),t(y)]=0\,, 
\label{commutatitivity}
\end{equation}
which implies that the corresponding eigenvalues 
$\Lambda(x)$ have the form
\begin{equation}
    \Lambda(x) = \frac{S(x)}{(1+ax)(1+bx)} \,,
    \label{Lambda}
\end{equation}    
where $S(x)$ is a polynomial in $x$ of order $L+2$. (The factors in the 
denominator in (\ref{Lambda}) are due to corresponding factors in 
the definitions of $\widetilde K(x)$ and $K(x)$ (\ref{RKmats}).)
The
proof of (\ref{commutatitivity}) given in \cite{skl} does not apply 
here since $R^{t_1}$ is not invertible; nevertheless, an alternative 
proof is available \cite{CRV}.

The deformed Markovian matrix \eqref{eq:Mmu} is recovered from the 
transfer matrix (\ref{transfer}) as follows
\begin{equation}
 M(\mu)= -\frac{1}{2} \frac{d}{dx} t(x)\Big|_{x=1}\;.\label{eq:Mmut}
\end{equation}
Therefore, once we know the eigenvalues of the 
transfer matrix, we readily obtain the eigenvalues of the 
deformed Markovian matrix,
\begin{equation}
\lambda(\mu) = -\frac{1}{2} 
\frac{d}{dx} \Lambda(x)\Big|_{x=1} \,.
\label{lambdaLambda}
\end{equation}
The higher charges defined by
\begin{equation}
H^{(r)}=\frac{(-1)^r}{2(r-1)!} \frac{d^r}{dx^r} \ln\left( t(x) 
\right) \ \Big|_{x=1}\,, \qquad r = 1, 2, \ldots  
\label{highercharges}
\end{equation}
commute among themselves as a further consequence of the 
commutativity property (\ref{commutatitivity}). Since $t(1)=1$, we see that $H^{(1)}= 
M(\mu)$.

In \cite{C14}, the eigenvalues of the transfer matrix $t(x)$ have been obtained by using the modified algebraic ansatz \cite{BC}:
\begin{equation}\label{eq:Lam}
 \Lambda(x)=x^{L+1} \frac{b+x}{bx+1} 
 \prod_{k=1}^L\frac{xu_k-1}{u_k-x} 
 -\frac{(x-1)^{2L}(ax+e^\mu)(x^2-1)}{(ax+1)(bx+1)}\prod_{k=1}^L\frac{u_k}{u_k-x}\,,
\end{equation}
where the parameters $\{\ u_k\ |\ k=1,2,\dots,L\}$, called Bethe roots, must satisfy the following Bethe equations
\begin{equation}
 (a u_j +e^\mu)\left( u_j-1 \right)^{2L}(u_j^2-1)=u_j^{L+1}(au_j+1)(u_j+b)\prod_{k=1}^L\left(u_j-\frac{1}{u_k}\right)\quad\text{for}\quad j=1,2,\dots,L\;.\label{eq:BEu}
\end{equation}
The eigenvectors are also computed in \cite{C14}, but in this letter we focus only on the eigenvalues.

Let us introduce the following polynomial w.r.t. the spectral parameter 
\begin{equation}
 Q(x)=\prod_{k=1}^L \left(1-\frac{x}{u_k}\right)\;,
\end{equation}
whose the zeros are the Bethe roots. 
This polynomial is usually called the $Q$-polynomial and allows us to rewrite relation \eqref{eq:Lam} for the eigenvalues as follows
\begin{equation}\label{eq:TQ}
\Lambda(x)\, Q(x)=x^{2L+1} \frac{b+x}{bx+1}Q(1/x) -\frac{(x-1)^{2L}(ax+e^\mu)(x^2-1)}{(ax+1)(bx+1)}\;. 
\end{equation}
This relation is called the (functional) T-Q relation. 
 
Although the transfer matrix (\ref{transfer}) has $2^{L}$ eigenvalues, we shall 
henceforth restrict our attention to only one of them, namely, the 
(unique) eigenvalue that tends to $1$ as $\mu$ tends to $0$
\begin{equation}
    \Lambda(x)\Big\vert_{\mu=0} = 1 \,.
    \label{zeromulimit}
\end{equation}   
We are interested in this eigenvalue because it corresponds to the eigenvalue $\lambda(\mu)$ of 
the deformed Markovian matrix $M(\mu)$ (\ref{eq:Mmu}) with the 
largest real part, which is the 
only eigenvalue of $M(\mu)$ that tends to $0$ as $\mu$ tends to $0$,
see (\ref{lambdaLambda}).

\subsection{Expansion for small $\mu$ \label{sec:smmu}}

The fact (\ref{zeromulimit}) that $\Lambda(x)$ has a simple limit for 
$\mu=0$ suggests to solve the T-Q relation perturbatively in terms of $\mu$.
This idea has been exploited previously for the periodic exclusion processes \cite{PM,Pro}. 

\begin{figure}[htb]
 \centering 
 \includegraphics[scale=0.3]{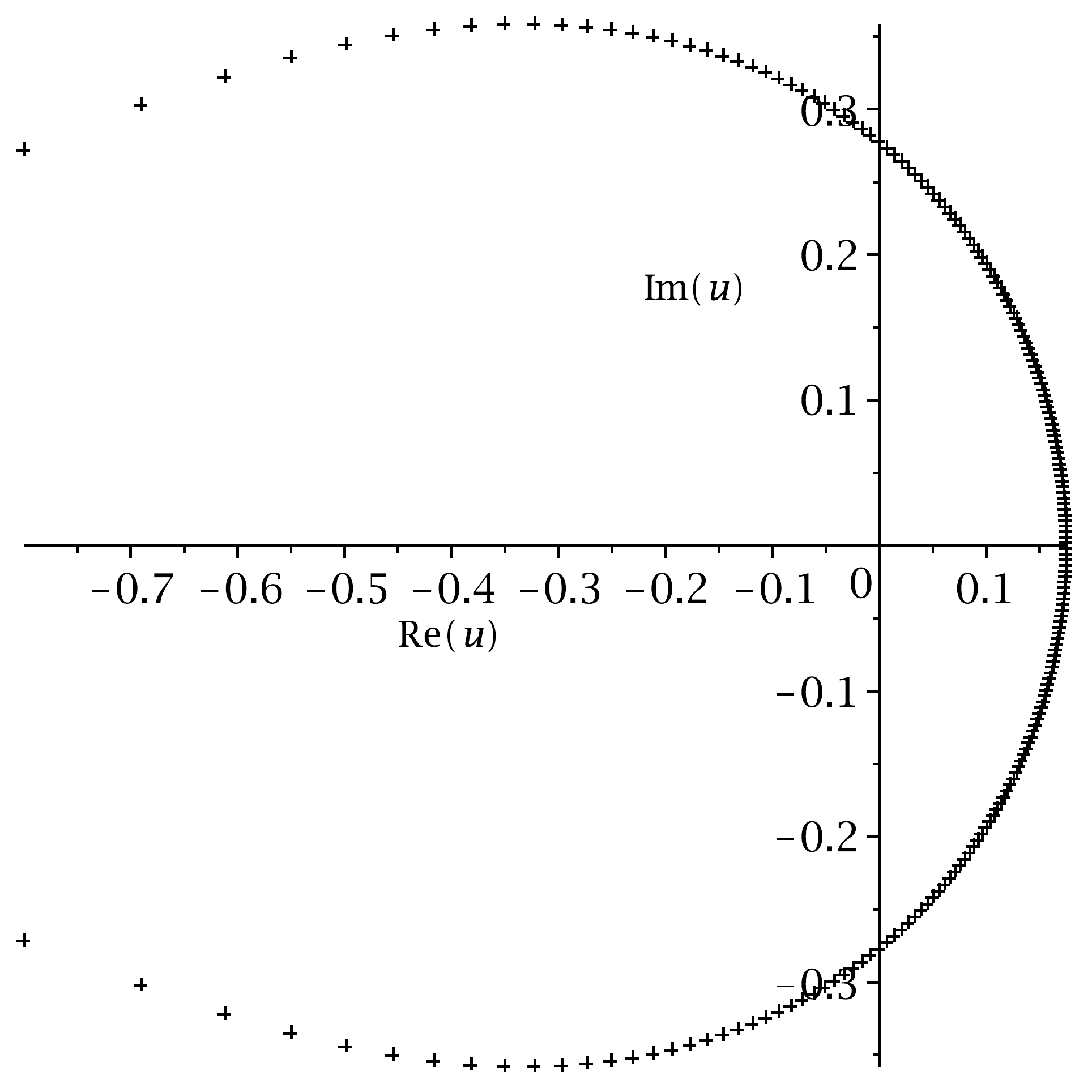}
 \caption{Bethe roots of \eqref{eq:BEu}  for $\mu=0$, $L=200$ and $b=0.8$.\label{fig:exbethe}}
\end{figure}

Let us introduce the following expansions, for small $\mu$, 
\begin{eqnarray}
 \Lambda(x)&=&1+\mu\, \Lambda^{(1)}(x) +\mu^2\, \Lambda^{(2)}(x) +\dots \label{eq:exL}  \\
 Q(x)&=&Q^{(0)}(x)+\mu\, Q^{(1)}(x) +\mu^2\, Q^{(2)}(x) + \dots \label{eq:exQ}
\end{eqnarray}
where
\begin{eqnarray}
&& Q^{(0)}(x)=\sum_{k=0}^L q^{(0)}_k x^k\quad,\qquad Q^{(j)}(x)=\sum_{k=1}^L q^{(j)}_k x^k \;, \label{eq:Qx}\\
&& \Lambda^{(j)}(x)=\frac{1}{(ax+1)(bx+1)}\sum_{k=0}^{L+2} \ell^{(j)}_k x^k\;.\label{eq:Lx}
\end{eqnarray}
We can easily show from the T-Q relation that $\Lambda(0)=e^\mu$, which implies $\ell^{(j)}_0=\frac{1}{j!}$.
The T-Q relation \eqref{eq:TQ} implies
\begin{eqnarray}
&& Q^{(0)}(x)=x^{2L+1} \frac{b+x}{bx+1}Q^{(0)}(1/x) -\frac{(x-1)^{2L}(x^2-1)}{bx+1}\;,\label{eq:TQdeg0}\\
 &&Q^{(1)}(x)+Q^{(0)}(x)\Lambda^{(1)}(x)=x^{2L+1} \frac{b+x}{bx+1}Q^{(1)}(1/x) -\frac{(x-1)^{2L}(x^2-1)}{(ax+1)(bx+1)}\;,...\label{eq:TQdeg1}
\end{eqnarray}
and so on. 
With these choices of expansions, these equations have a unique 
solution, which corresponds to the largest eigenvalue $\lambda(\mu)$ since one gets $\Lambda(x) \xrightarrow[\mu\rightarrow 0]{} 1$.
In particular, we can find a closed formula for $Q^{(0)}(x)$:
\begin{eqnarray}
 Q^{(0)}(x)= 2\sum_{k=0}^L  (-x)^k \sum_{p=0}^k  b^{k-p}\  \frac{L-p+1}{2L-p+2}   \begin{pmatrix}
                                           2L+1\\ p
                                          \end{pmatrix}\;.\label{eq:Q0L1}
\end{eqnarray}
Let us emphasize that $Q^{(0)}(x)$ corresponds to the $Q$-polynomial
for the non-deformed ($\mu=0$) Markovian model, and its $L$ roots are
the Bethe roots solution of the Bethe equations \eqref{eq:BEu} for
$\mu=0$, corresponding to the eigenvalue $\Lambda(x)=1$.  Using
this explicit formula for the $Q$-polynomial, we can easily find the Bethe
roots for very large systems.  We display an example in Figure \ref{fig:exbethe}.
Let us recall that a surprising connection between the $Q$-polynomial
$Q^{(0)}(x)$ and the normalization of the stationary state has been
discovered in \cite{CMRV}, which relates the roots of the
$Q$-polynomial $Q^{(0)}(x)$ and the Lee-Yang zeros studied in
\cite{BE}.

\section{Equivalent T-Q relation and exact eigenvalues of the higher charges}

\subsection{Equivalent T-Q relation \label{sec:eqTQ}}

In the previous section, we reviewed
the T-Q relation for the open TASEP.  We then explained how to expand the 
eigenvalue with the property (\ref{zeromulimit})
for small $\mu$.
Remarkably, there exists another T-Q relation which gives exactly 
the same eigenvalue $\Lambda(x)$, for any value of $L$. Indeed, 
the solution $\Lambda(x)$ of the T-Q relation \eqref{eq:TQ} with the 
expansions \eqref{eq:exL}-\eqref{eq:Lx}, is also a solution of the following T-Q relation
\begin{equation}\label{eq:TQn}
 \Lambda(x)\, \overline Q(x) = x^{L}(x+b)(x+a) e^{-\mu} +\overline Q(0) \frac{(1-x)^{2L+2}(x+1)^2}{(ax+1)(bx+1)} e^\mu\;,
\end{equation}
where the expansion of $\overline Q(x)$ is given by
\begin{eqnarray}
 &&\overline Q(x)=\overline Q^{(0)}(x)+\mu\, \overline Q^{(1)}(x) 
 +\mu^2\, \overline Q^{(2)}(x) +\dots\\
 &&\overline Q^{(0)}(x)=\sum_{k=0}^{L+2} \overline{q}^{(0)}_k x^k\quad,\qquad \overline Q^{(j)}(x)=\sum_{k=0}^{L+1} \overline{q}^{(j)}_k x^k\;. \label{eq:exQb}
\end{eqnarray}
If we add the additional requirement
$\ell^{(j)}_0=\frac{1}{j!}$ (see below (\ref{eq:Lx})), then the solution of the T-Q relation \eqref{eq:TQn} is unique.
We did not succeed in proving the equivalence between the T-Q 
relations (\ref{eq:TQ}) and  (\ref{eq:TQn}); however,
we have strong evidence that this equivalence holds. For $a=b=0$, we get the same 
expansions of $\Lambda(x)$ up to the order $\mu^{30}$ and for $L=1,2,\dots, 35$ from both T-Q relations. Similar positive results have been obtained for $a\neq 0$ and $b\neq 0$, 
up to the order $\mu^5$ and $L=1,\dots,5$. Moreover, 
both T-Q relations with $a=b=0$ have the same exact solution $\Lambda(x)$ for 
$L=0$ and $L=1$
\begin{align}
    L &=0:  
    %&\Lambda(x) &= x^{2} - e^{\mu}(x^{2}-1)
    &\Lambda(x) &= 1 -  (e^{\mu}-1)(x^{2}-1)    
    \,, \qquad Q(x) = 1 
    \,, \qquad \overline Q(x)= x^{2} +e^{-\mu} -1  \,, \label{L0exact}\\
     L &=1:  
     %&\Lambda(x) &= x^{2} + e^{\mu}(x^{2}-1)(x-1) - e^{\mu/2}  x (x^{2}-1)  \,, \non\\
     &\Lambda(x) &= 1+ (e^{\mu/2}-1)(e^{\mu/2}x-e^{\mu/2}-1)(x^{2}-1)  \,, \non\\
    &  & Q(x) &= 1 - x - e^{-\mu/2} x \,, \non\\
     & & \overline Q(x) &= (x^{2}-1)(x-1) + e^{-\mu} x +  e^{-\mu/2} 
     (x^{2}-1) \,.  \label{L1exact}
\end{align}    

Let us emphasize that, although $\Lambda(x)$ is the same for both T-Q relations, the $Q$-polynomials $Q(x)$ and $\overline Q(x)$ are different. 
For example, one can see that
\begin{equation}
 \overline Q^{(0)}(x)=x^L(x+a)(x+b) \,,
 \label{eq:Q0b}
\end{equation}
which is different from $Q^{(0)}(x)$ given by \eqref{eq:Q0L1}. 
See also (\ref{L0exact}) and  (\ref{L1exact}).

As usual, we call the roots of $\overline Q(x)$ Bethe roots, and we denote them by $\overline u_j$. Due to the expansion \eqref{eq:exQb} and 
of the explicit form \eqref{eq:Q0b} of $\overline Q^{(0)}(x)$, we know that 
\begin{equation}
 \overline Q(x)=\prod_{k=1}^{L+2}(x-\overline u_k)\;.\label{eq:Qub}
\end{equation}
By knowing the explicit expression \eqref{eq:Q0b} of $\overline
Q^{(0)}(x)$, we deduce that $L$ Bethe roots $\overline u_k$ tends to
$0$ when $\mu\rightarrow 0$, one tends to $-a$, and the last one 
tends to
$-b$.

From the T-Q relation \eqref{eq:TQn}, we deduce that these Bethe roots satisfy the following Bethe equations, for $j=1,2,\dots, L+2$,
\begin{equation}\label{eq:BEb}
\overline u_j^{L}(\overline u_j+b)(\overline u_j+a)(a\overline u_j+1)(b\overline u_j+1) =(-1)^{L+1} e^{2\mu}\ (1-\overline u_j)^{2L+2}(\overline u_j+1)^2 \prod_{k=1}^{L+2}\overline u_k\;.
\end{equation}

At this point, the new T-Q relation \eqref{eq:TQn} seems artificial. However, for $a=b=0$, it becomes 
\begin{equation}\label{eq:TQnn}
 \Lambda(x)\, \overline Q(x) = x^{L+2} e^{-\mu} +\overline Q(0) (1-x)^{2L+2}(x+1)^2 e^\mu\;,
\end{equation}
and we recognize the T-Q relation associated to the diagonalization of the following transfer matrix
\begin{equation}
 \bar t(x)=\tr_0\left(\ Z_0 R_{0,2L+4}(-x) R_{0,2L+3}(-x) 
 R_{0,2L+2}(x) R_{0,2L+1}(x)\dots R_{01}(x)\ \right)\;,
\end{equation}
where $Z=\diag( e^{-\mu}\ , \ e^\mu\ )$. 
The matrix $\bar t(x)$ is evidently a transfer matrix for a TASEP with $2L+4$ sites, 
quasi-periodic boundary conditions (with twist matrix $Z$), and 
inhomogeneities at sites $2L+4$ and $2L+3$ (responsible for the minus 
signs in the corresponding $R$-matrices).
The T-Q relation \eqref{eq:TQnn} corresponds to the sector with $L+2$ particles.

We note that the T-Q relation \eqref{eq:TQn} may be obtained as a limit of the T-Q relation obtained in \cite{LP} by functionnal Bethe ansatz for the ASEP.

\subsection{Exact expansions for the eigenvalues of the higher charges \label{sec:ex}}

Since the T-Q relation \eqref{eq:TQn} is closely related to the one for the TASEP with periodic boundary conditions, 
the computations done previously for the periodic case can be 
generalized to obtain results for the T-Q relation \eqref{eq:TQn} and, by consequence, 
for the TASEP with open boundaries. In particular, we want to 
generalize the results of \cite{DL} to obtain exact 
expansions for the eigenvalues of the higher conserved charges (\ref{highercharges}).

\paragraph{Case with $\boldsymbol{a=b=0}$.}
By setting $a=b=0$ and by performing the change of variables $\overline u_j=1-\frac{z_j}{\Delta}$ with $\Delta^{L+2}=e^\mu$, the Bethe equations \eqref{eq:BEb} become, for $j=1,2,\dots, L+2$,
\begin{equation}\label{eq:Dzj}
 (\Delta-z_j)^{L+2}+ z_j^{2L+2}(2\Delta-z_j)^2 \prod_{k=1}^{L+2} (z_k-\Delta)=0\;.
\end{equation}
In this case, all the Bethe roots $z_j$ tend to $1$ when $\mu\rightarrow 0$.

From the T-Q relation \eqref{eq:TQ}, we deduce also that
$\Lambda(1)=1$.  By setting this result in the T-Q relation
\eqref{eq:TQnn}, and by using the explicit expression \eqref{eq:Qub} of
$\overline{Q}(x)$, one obtains
\begin{equation}
 \prod_{k=1}^{L+2} z_k =1\;.\label{eq:zz1}
\end{equation}
In \cite{DL}, the authors succeeded in computing a development for the largest eigenvalue for the periodic TASEP starting from relations similar to \eqref{eq:Dzj} and \eqref{eq:zz1}. 
We shall adapt their approach to compute the eigenvalues 
of the first $2L+1$ conserved charges (\ref{highercharges})
\begin{eqnarray}
 I^{(r)}&=& \frac{(-1)^r}{2(r-1)!} \frac{d^r}{dx^r} \ln\left( \Lambda(x) \right) \ \Big|_{x=1}\quad\text{for}\quad r=1,2,\dots,2L+1\\
 &=& \sum_{j=1}^{L+2}h^{(r)}(z_j)\quad \text{where}\quad h^{(r)}(z)=\frac{1}{2}\left(\frac{\Delta^r-z^r}{z^r}\right)\;.\label{eq:Ir}
\end{eqnarray}
The equality \eqref{eq:Ir} has been obtained by using the T-Q relation \eqref{eq:TQnn}. 
Note that $I^{(1)}$ corresponds to the largest eigenvalue for the deformed Markovian matrix.

Let us introduce the following polynomial 
\begin{equation}
 P(z)=(\Delta-z)^{L+2}-Az^{2L+2}(2\Delta-z)^2\;.
\end{equation}
Then, by Cauchy's theorem, we obtain the following expression for the 
eigenvalues 
\begin{equation}\label{eq:Sh}
 I^{(r)}=\sum_{j=1}^{L+2}h^{(r)}(z_j)= \oint_{|z-\Delta|=\epsilon} \frac{dz}{2i\pi}\ h^{(r)}(z) \frac{P'(z)}{P(z)}\;,
\end{equation}
where $z_j$ are the $L+2$ solutions of $P(z)=0$ satisfying $z_j\xrightarrow[A\rightarrow 0]{} \Delta$.
By using the explicit form of $P(z)$, this expression becomes
 \begin{equation}\label{eq:Shex}
 I^{(r)}= -\oint_{|z-\Delta|=\epsilon} \frac{dz}{2i\pi}\ \frac{(\Delta^r-z^r)(\Delta-z)^{L+1}}{2 z^{r+1}(2\Delta-z)}\ \frac{(L+2)z^2+4\Delta(L+1)(\Delta-z)}{(\Delta-z)^{L+2}-Az^{2L+2}(2\Delta-z)^2}\;.
\end{equation}
We have used the fact that $P(z_j)=0$ to simplify the $A$ in the numerator.
 
By expanding w.r.t. $A$ in the 
integrand of \eqref{eq:Shex} and 
performing the integrals, we eventually obtain, for $r=1,2,\dots,2L+1$, 
\begin{equation}
 I^{(r)}= -\frac{r}{2} \sum_{k=1}^{\infty} B^k \frac{(2k)! (2k(L+1)-r-1)!}{k(k(L+2)-1)!} \sum_{p=0}^{\left[\frac{r-1}{2}\right]}  \begin{pmatrix}
                                                                                                                                             r-1\\2p
                                                                                                                                            \end{pmatrix}
\frac{(-1)^p}{(k-p)!(k(L+1)-r+p)!}
\end{equation}
with $B=-A(-\Delta)^{L+2}$. In particular, for $r=1$, 
 \begin{equation}
   I^{(1)}= -\frac{1}{2} \sum_{k=1}^{\infty} B^k \frac{(2k)! (2k(L+1)-2)!}{k(k(L+2)-1)!}  \
\frac{1}{k!(k(L+1)-1)!}\;.\label{eq:I1}
 \end{equation}

Similarly, with the help of \eqref{eq:zz1}, one gets
\begin{equation}\label{eq:muh}
 0=\sum_{j=1}^{L+2}\ln(z_j)= -\oint_{|z-\Delta|=\epsilon} \frac{dz}{2i\pi}\ \frac{\ln(z)(\Delta-z)^{L+1}}{ z(2\Delta-z)}\ \frac{(L+2)z^2+4\Delta(L+1)(\Delta-z)}{(\Delta-z)^{L+2}-Az^{2L+2}(2\Delta-z)^2}
\end{equation}
and, by expanding w.r.t. $A$ as before, we obtain
\begin{equation}
 \mu= -\sum_{k=1}^{\infty} B^k \frac{(2k)!\; (2k(L+1))!}{2k\; (k(L+1))!\; (k(L+2))!\; k!}\;.\label{eq:muB}
\end{equation}
Expansions \eqref{eq:I1} and \eqref{eq:muB} of $I^{(1)}$ and $\mu$ in 
terms of $B$ reproduce exactly the matrix ansatz result of \cite{LM}.
We have generalized their result by computing also 
expansions for the eigenvalues $I^{(r)}$ of the higher conserved charges.

\paragraph{Case with $\boldsymbol{a\neq0}$ and $\boldsymbol{b\neq 
0}$.} This case can be treated as previously. The main difference is 
that $L$ Bethe roots $z_j$ tend to $1$, while the last two Bethe roots 
tend to $1+a$ and $1+b$. We must take in account this difference when 
we define the contour of integration. 
Then, the previous computation can be done for this more general case 
by using instead the following polynomial
\begin{equation}
 P(z)=(\Delta-z)^{L}(\Delta(1+a)-z)(\Delta(1+b)-z)(\Delta(1+a)-az)(\Delta(1+b)-bz) -A\Delta^2 z^{2L+2}(2\Delta-z)^2\;,
\end{equation}
in view of the Bethe equations (\ref{eq:BEb}).
However, the integrals are significantly more complicated, and we 
have not managed to obtain comparably simple results for general values of $r$.

\paragraph{Acknowledgements: } 
We dedicate this paper to the memory of Vladimir Rittenberg, who greatly influenced our work and was also a dear friend.

N.~Cramp\'e acknowledges the 
University of Miami where this work was done, and thanks M.~Vanicat for helpful preliminary discussions.

\end{document}